\documentstyle[11pt,appb,epsf]{article} 

\begin{document}
\null
\hfill WUE-ITP-97.045\\
\null
\hfill hep-ph/9712354\\
\vskip .8cm
\begin{center}
{\Large \bf Angular and Energy Distribution in \\[.5em]
Neutralino 
 Production 
 and Decay with \\[.5em]
complete Spin Correlations%
\footnote{Work supported by the German Federal Ministry for
Research and Technology (BMBF) under contract number
05 7WZ91P (0).}
}
\vskip 2.5em
{\large
{\sc G. Moortgat-Pick}%
\footnote{e-mail: gudi@physik.uni-wuerzburg.de}
}\\[1ex]
{\normalsize \it
Institut f\"ur Theoretische Physik, Universit\"at W\"urzburg, Am
Hubland, D-97074~W\"urzburg, Germany}\\[2ex]
{\large
{\sc H. Fraas}%
\footnote{e-mail: fraas@physik.uni-wuerzburg.de}
}\\[1ex]
{\normalsize \it Institut f\"ur Theoretische Physik, Universit\"at
W\"urzburg,\\
D--97074~W\"urzburg, Germany}
\vskip 2em
\end{center} \par
\vfil
{\bf Abstract} \par
We study the process $e^{-}e^{+}\to\tilde{\chi}^0_1\tilde{\chi}^0_2$ 
 with the subsequent
decay $\tilde{\chi}^0_2 \to \tilde{\chi}^0_1 \ell^{+}\ell^{-}$ 
 taking into account the complete spin correlations between production
and decay. 
We present numerical results for the
 lepton angular distribution and the
distribution of the opening angle between the outgoing leptons 
 for $\sqrt{s}=182$ GeV. We examine 
 representative mixing scenarios in the MSSM and  
study the influence of
 the common scalar mass parameter $m_0$.
 For the lepton angular distribution the effect of the 
spin correlations amounts to up to 20\%.
 The shape of the lepton angular distribution is very
sensitive to the mixing in the gaugino sector and to the value
of $m_0$. We find that the opening angle distribution is 
 suitable for distinguishing between Higgsino-like and gaugino-like
neutralinos.
\par
\vskip 1cm
\null
\setcounter{page}{0}
\clearpage


\section{Introduction}
The most economical candidate for a realistic SUSY model with mini\-mal 
content of particles
is the Minimal Supersymmetric Extension of the Standard Model (MSSM).
 Here the SUSY particles can only be produced
in pairs and the lightest supersymmetric particle (LSP), usually
assumed to be the lightest neutralino
$\tilde{\chi}^0_1$, is stable and
escapes detection.

Angular
 distributions and angular correlations of the decay products of
neutralinos can give
valuable information on their mixing character and are necessary for
constraining the parameter space of the MSSM.
 Since decay angular distributions depend on the polarization of the
parent particles one has to take into account
spin correlations between production and decay. 
\section{General Formalism}
 Both the production process, 
$e^{-} e^{+} \to \tilde{\chi}^0_1 \tilde{\chi}^0_2$, and
the decay process, $\tilde{\chi}^0_2 \to
\tilde{\chi}^0_1 \ell^{+} \ell^{-}$
 contain contributions from $Z^0$ exchange in the
direct channel and from $\tilde{\ell}_L$ and $\tilde{\ell}_R$
 exchange in the crossed channels [1].

The helicity of the decaying neutralino $\tilde{\chi}^0_2$ is denoted
by $\lambda_2$. 
 The amplitude $T=\Delta_2 P^{\lambda_2} D_{\lambda_2}$
 of the combined process is
a sum over all polarization states of the helicity amplitude
$P^{\lambda_2}$ for the production process times the helicity amplitude
 $D_{\lambda_2}$ for the decay process and a pseudopropagator
$\Delta_2=1/[s_2-m_2^2+i m_2 \Gamma_2]^{-1}$ of $\tilde{\chi}^0_2$.
The effective mass squared of $\tilde{\chi}^0_2$ is denoted by $s_2$, the mass
by $m_2$ and the width by $\Gamma_2$.
The amplitude squared 
$|T|^2=|\Delta_2|^2 \rho^{\lambda_2 \lambda_2'}_P
\rho^D_{\lambda_2' \lambda_2} \label{N}$
 is thus composed from the
unnormalized spin density matrix
 $\rho^{\lambda_2 \lambda_2'}_P=P^{\lambda_2} P^{\lambda_2' *}$
of $\tilde{\chi}^0_2$ and
 the decay matrix
 $\rho^{D}_{\lambda_2' \lambda_2}=D_{\lambda_2} D_{\lambda_2'}^{*}$
for the respective decay channel.
 All helicity indices but that of the decaying neutralino are
 suppressed. Repeated indices are summed over [2].

Interference terms between
various helicity amplitudes preclude factorization in a production
factor  $\sum_{\lambda_2} |P^{\lambda_2}|^2$ times a decay factor
$\overline{\sum}_{\lambda_2} |D_{\lambda_2}|^2$ as for the case of spinless
particles.

\vspace{.3cm}
We split the phase space into that for production,  
and that for decay and obtain
the differential cross section in the $e^{-}$-$e^{+}$-cms by integrating
over the effective mass squared $s_2$ of  
$\tilde{\chi}^0_2$. 
 Since the total width of 
 $\tilde{\chi}^0_2$ is much smaller than its mass 
($\Gamma_2\ll m_2$)
we make the narrow width approximation: 
$ \frac{1}{(s_2-m_2^2)^2+m_2^2 \Gamma_2^2}
\approx \frac{\pi}{m_2 \Gamma_2} \delta(s_2-m_2^2)$. 

\section{Numerical Results and Discussion}
Neutralinos are linear superpositions $(\tilde{\gamma}|\tilde{Z}|\tilde{H}^0_a|
\tilde{H}^0_b)$
 of the photino $\tilde{\gamma}$ and 
the zino $\tilde{Z}$, coupling to sleptons, and the two 
Higgsinos $\tilde{H}^0_a$ and $\tilde{H}^0_b$, coupling to $Z^0$. 
The composition of the neutralino states depend on the three
SUSY mass parameters $M, M'$ ( with $M'=\frac{5}{3} M \tan^2\theta_W$ suggested
by GUT [3])
 and $\mu$, and on the ratio $\tan\beta=v_2/v_1$ of the vacuum expectation
values of the Higgs fields. We choose $\tan\beta=2$.  
 The masses of the sleptons are determined by $M$ and
 $\tan\beta$ and the common
scalar mass parameter $m_0$ [4]. 
 We shall consider three representative scenarios which differ significantly
in the nature of the two lowest mass eigenstates $\tilde{\chi}^0_1$
and $\tilde{\chi}^0_2$ ( (A) M=78 GeV, $\mu=-250$ GeV; 
 (B) M=100 GeV, $\mu=400$ GeV; (C) M=210 GeV, $\mu=-60$ GeV )
 and two values of the scalar mass,
 $m_0=80$ GeV and $m_0=200$ GeV. 
In (A) $\tilde{\chi}^0_1=(+.94|-.32|-.08|-.07)$ 
has a dominating photino component and
$\tilde{\chi}^0_2=(-.34|-.90|-.16|-.23)$ has a dominating zino
component. In
 (B) both neutralinos, $\tilde{\chi}^0_1=(-.76|+.63|-.13|+.09)$
and $\tilde{\chi}^0_2=(+.65|+.73|-.18|-.10)$, 
are nearly equal photino-zino mixtures.
 In (C) both neutralino states are dominated by strong
Higgsino components, $\tilde{\chi}^0_1=(-.10|+.17|-.19|+.96)$ and 
$\tilde{\chi}^0_2=(+.06|-.31|+.92|+.24)$.
 
  The total cross section for the combined process is 
independent of the spin correlations [5].
 For $\sqrt{s}=182$ GeV and $m_0=80$ GeV ($m_0=200$ GeV) we obtain in
 (A) 36.1 fb (11.1 fb), in (B) 15.4 fb (2.9 fb) and in (C) 57.1 fb (57.2 fb). 
\subsection{Lepton angular distributions}
We calculate the distribution of the lepton angle between
the incoming $e^{-}$ and the outgoing lepton $\ell^{-}$ in the
laboratory system, $d\sigma/d\cos\Theta_{-}$, and 
 compare our results
with those obtained from the assumption of factorization of the
differential cross section into production and decay (Figs.~1--4).

The spin effect is sizeable in the gaugino-like scenario (A). 
 In the forward and backward direction it
 amounts to about $15\%$ for $m_0=80$ GeV
 and to more than
$20\%$ for $m_0=200$ GeV (Fig.~1).

 The magnitude 
of the resulting forward-backward (FB) asymmetry sensitively 
 depends on the mixing-character of the neutralinos.
 In the gaugino-like scenario (B) for both values of $m_0$ the 
contribution of spin correlations
only amounts to at most $2.5\%$ in the forward and backward direction. 
 In the case of gaugino-like neutralinos the
scalar mass $m_0$ crucially determines the shape of the angular
distributions.  
 For $m_0=80$ GeV it has a maximum nearly perpendicular to the beam
direction and is almost FB symmetric
(Fig.~2). 
 For $m_0=200$ GeV the shape has completely changed. 
It has a minimum in the backward hemisphere
 and the forward direction is favoured (Fig.~3). 

In the Higgsino-like scenario (C) both production and decay are
domi\-nated by $Z^0$-exchange. 
Therefore the dependence on $m_0$ is considerably
smaller and we give only numerical results for $m_0=80$ GeV.
Here the contribution of spin correlations is negligible, maximally 
$0.3\%$, 
 so that the angular distribution is practically FB-symmetric
(Fig.~4).
\subsection{The lepton opening angle distribution}
The distribution of the opening angle
between both outgoing leptons  
in the laboratory system, $d\sigma/d\cos\Theta_{+-}$,  
factorizes due to the Majorana character of
the neutralinos.

The distributions are similar for both
gaugino-like scenarios (A) and (B) (Figs.~5 and 6).
 In (B) they are symmetric with a maximum at $\Theta_{+-}=\pi/2$
whilst for (A) larger angles between $\pi/2$ and $\pi$ are
favoured.

For Higgsino-like neutralinos (Fig.~7), 
 the shape is completely different from 
those of gaugino-like neutralinos.
Here the lepton pairs are preferably emitted with small angles
between them, approximately $60\%$ of them with
an opening angle between 0 and $\pi/2$.

The influence of varying the value of $m_0$ is rather small;
 the shape remains essentially unchanged.

It is obvious that the distribution of the opening angle between
the leptons is much more suitable for discrimination between
gaugino- and Higgsino-like neutralinos than the lepton 
angular distribution.
\subsection{Energy Distributions}
In all three mixing scenarios the energy
distribution of the outgoing lepton factorizes due to the Majorana
character of the neutralinos.

The shapes of all scenarios are similar, therefore we only show the
energy distribution of $\ell^{-}$ for (A) for $m_0=80$ GeV and
$m_0=200$ GeV (Fig.~8).
The maximum is independent of the actual value of $m_0$. For more
details see [6].
 As a consequence of CP invariance and the Majorana
character the energy spectra
 of both leptons, $\ell^{-}$ and $\ell^{+}$, are identical [7].
\section{Summary}
\vspace{-.3cm}
In this paper we have considered the associated production
 of neutralinos, $e^{-} + e^{+}\to
\tilde{\chi}^0_1 + \tilde{\chi}^0_2$,   
and the subsequent
direct leptonic decay, $\tilde{\chi}^0_2 \to 
\tilde{\chi}^0_1 +\ell^{+} + \ell^{-}$, 
 with complete spin correlations between production and decay.
 The quantum mechanical interference terms between the various
polarization states of the decaying neutralino give rise to a strong effect
in the lepton angular distribution,
 whereas the opening
angle distribution and the energy distribution are independent from these
spin correlations.

The opening angle distribution turns out to be suitable for distinguishing
 between Higgsino-like and gaugino-like neutralinos. However, it is 
 rather indifferent to variable mixing in the gaugino sector. The shape
of the opening angle distribution only slightly depends on the scalar mass
 $m_0$.
 The lepton angular distribution, on the other hand, is not only 
 very sensitive to the mixing in the gaugino sector but also to the
actual value of $m_0$.

G.~M.-P.\ thanks K.~Ko\l{}odziej and the other organizers of the Ustron
School for the friendly atmosphere during the Conference and the
excellent organization. We thank A.~Bartl and W.~Majerotto for many 
useful discussions. We are grateful to V.~Latussek for his support in 
the development of the numerical program.\\

\refname:

[1] A. Bartl, H. Fraas, W. Majerotto,
                 Nucl.Phys. {\bf B 278}, 1 (1986).

[2] H.E. Haber, in Proceedings of the 21st SLAC Summer Institute
                on 
 \hspace*{1.2cm}Particle Physics, Stanford 1993.

[3] H.E. Haber, G.L. Kane,  Phys. Rep. {\bf 117}, 75 (1985).

[4] L.J. Hall, J. Polchinski, Phys. Lett. {\bf B 152}, 335 
                 (1985).

[5] D.A. Dicus, E.C.G. Sudarshan, X. Tata, Phys. Lett. {\bf B 154}, 79
\hspace*{1.2cm}(1985).

[6] G. Moortgat-Pick, H. Fraas, hep-ph 9708481, for publ. in
                                Phys.Rev.{\bf D}.

[7] S.T. Petcov, Phys. Lett {\bf B139}, 421 (1984).

\newpage
\begin{picture}(10,5)
\put(0,0){\includegraphics{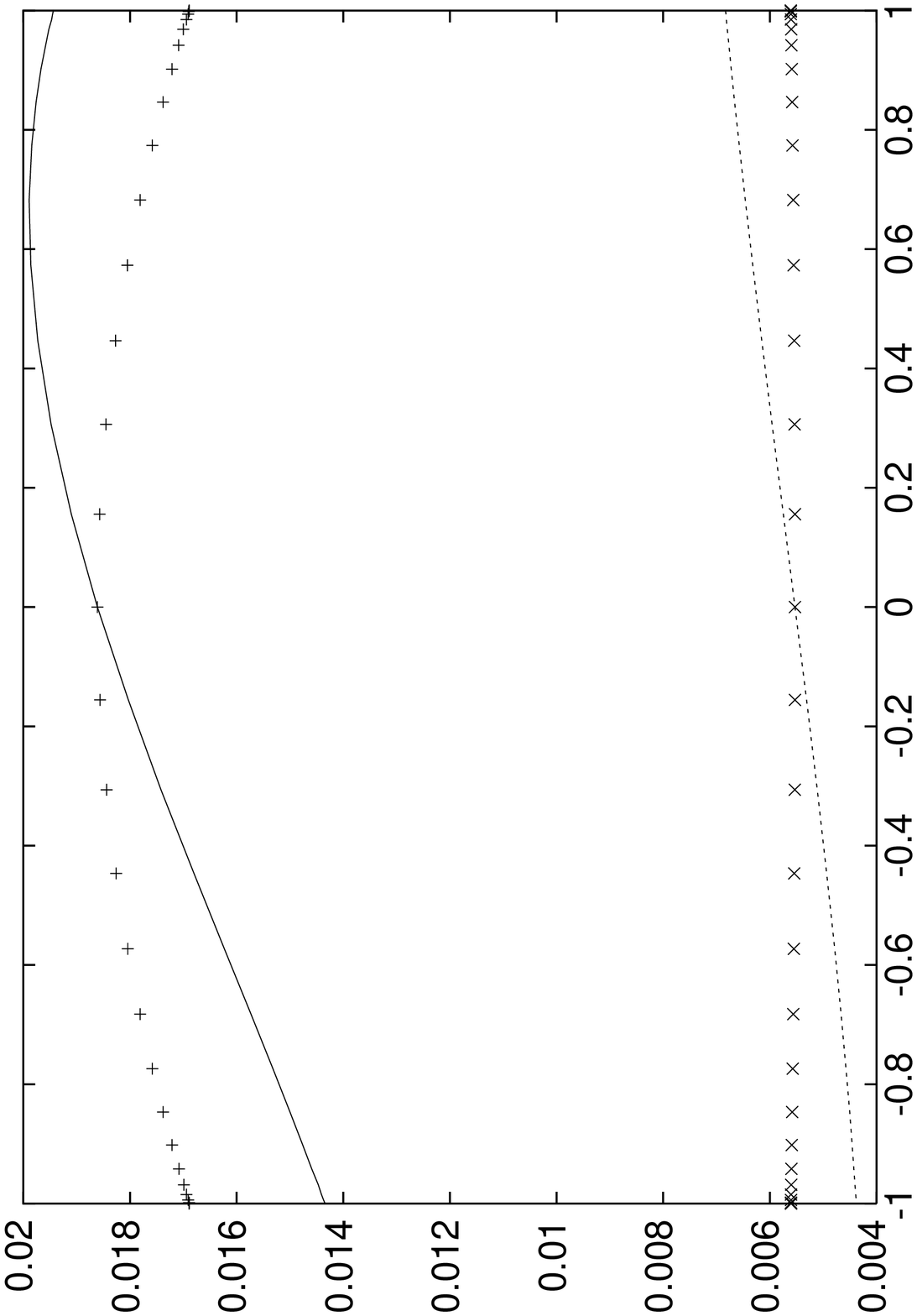}}
\put(130,-120){$ \scriptscriptstyle\cos\Theta_{-}$}
\put(-10,0){$ \scriptscriptstyle\frac{d\sigma}{d\Theta_{-}}/pb$}
\label{w1}
\put(-10,-156)
{\parbox{5.7cm}{\scriptsize Fig.~1: Lepton angular distribution in (A) for
for $m_0=80$ GeV with spin correlations fully
taken into account (upper solid) and
for assumed factorization (upper dotted); for $m_0=200$ GeV with spin
correlations (lower solid) and for assumed factorization (lower dotted).}}
\put(12,0){\includegraphics{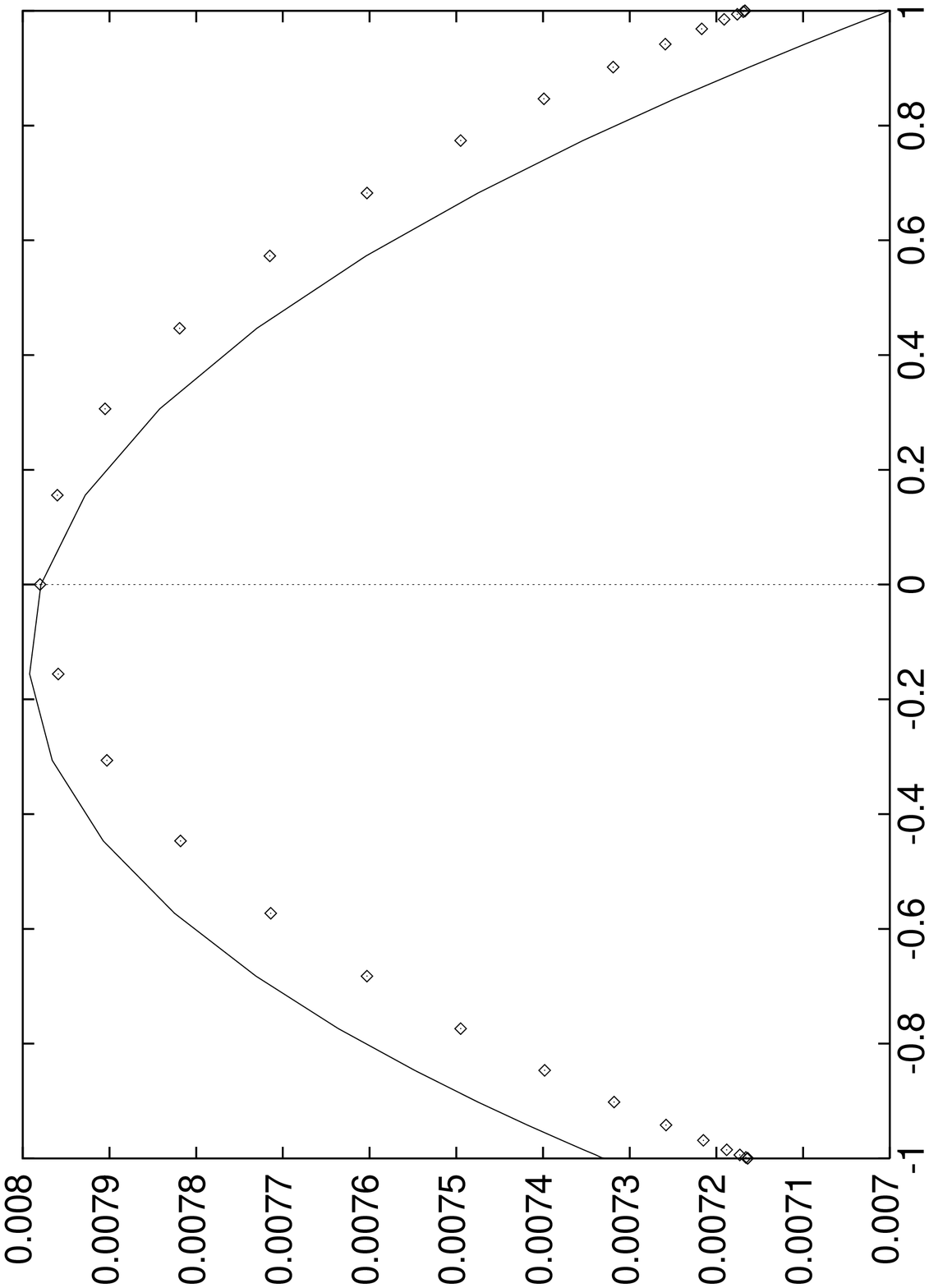}}
\put(310,-120){$\scriptscriptstyle\cos\Theta_{-}$}
\put(170,0){$ \scriptscriptstyle\frac{d\sigma}{d\Theta_{-}}/pb$}
\end{picture}
\label{80w2}
\put(170,-148)
{\parbox{5.7cm}{\scriptsize Fig.~2: Lepton angular distribution in (B) 
for $m_0=80$ GeV 
with spin correlations fully taken into account (solid) and for assumed
 factorization (dotted).}}

\vspace{.8cm}
\begin{picture}(10,5)
\put(0,0){\includegraphics{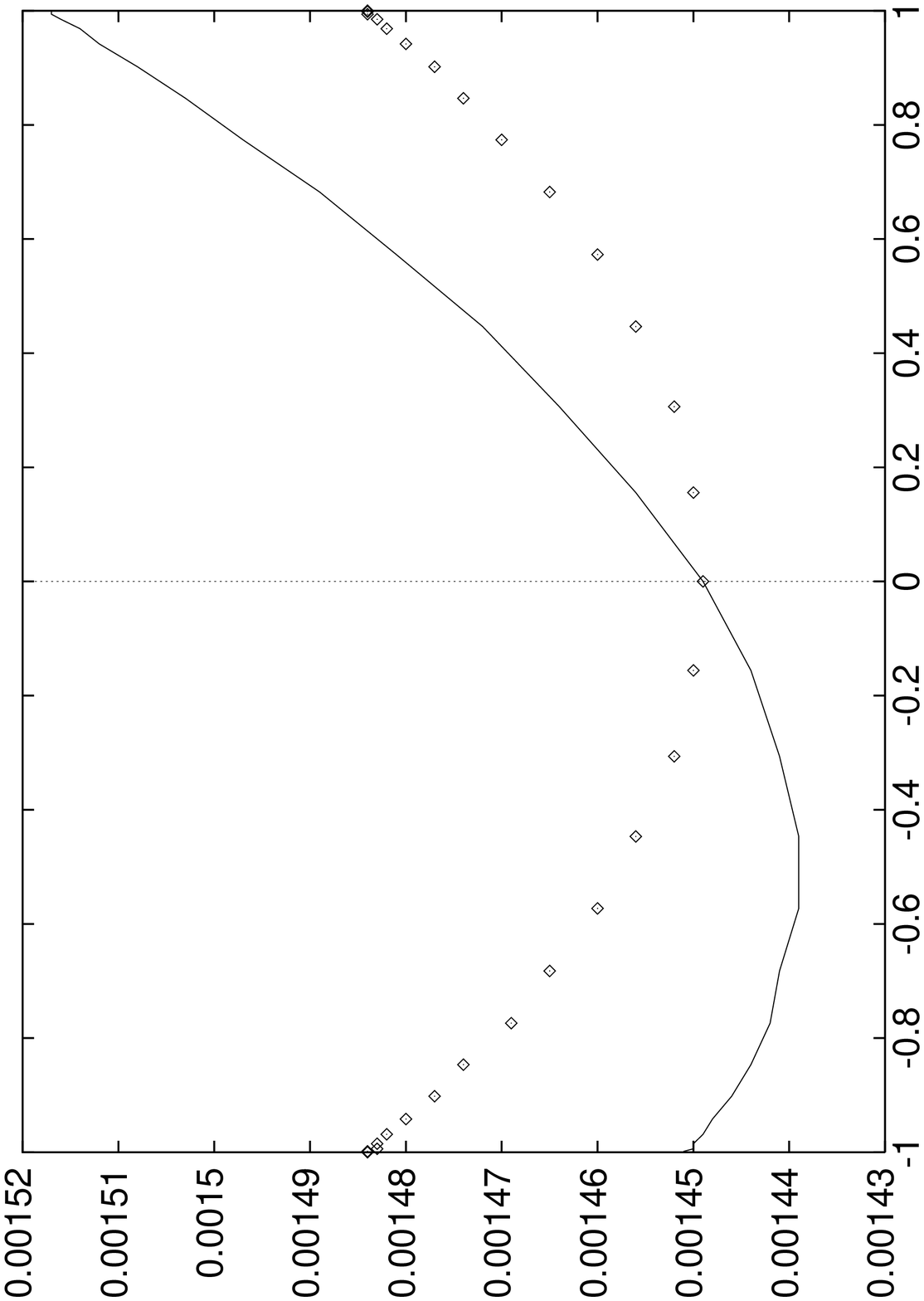}}
\put(130,-120){$\scriptscriptstyle \cos\Theta_{-}$}
\put(-10,0){$ \scriptscriptstyle \frac{d\sigma}{d\Theta_{-}}/pb$}
\label{200w2}
\put(-10,-148)
{\parbox{5.7cm}{\scriptsize Fig.~3: Lepton angular distribution in (B)
 for $m_0=200$ GeV
with spin correlations fully taken into account (solid)
and for assumed factorization (dotted).}}
\put(12,0){\includegraphics{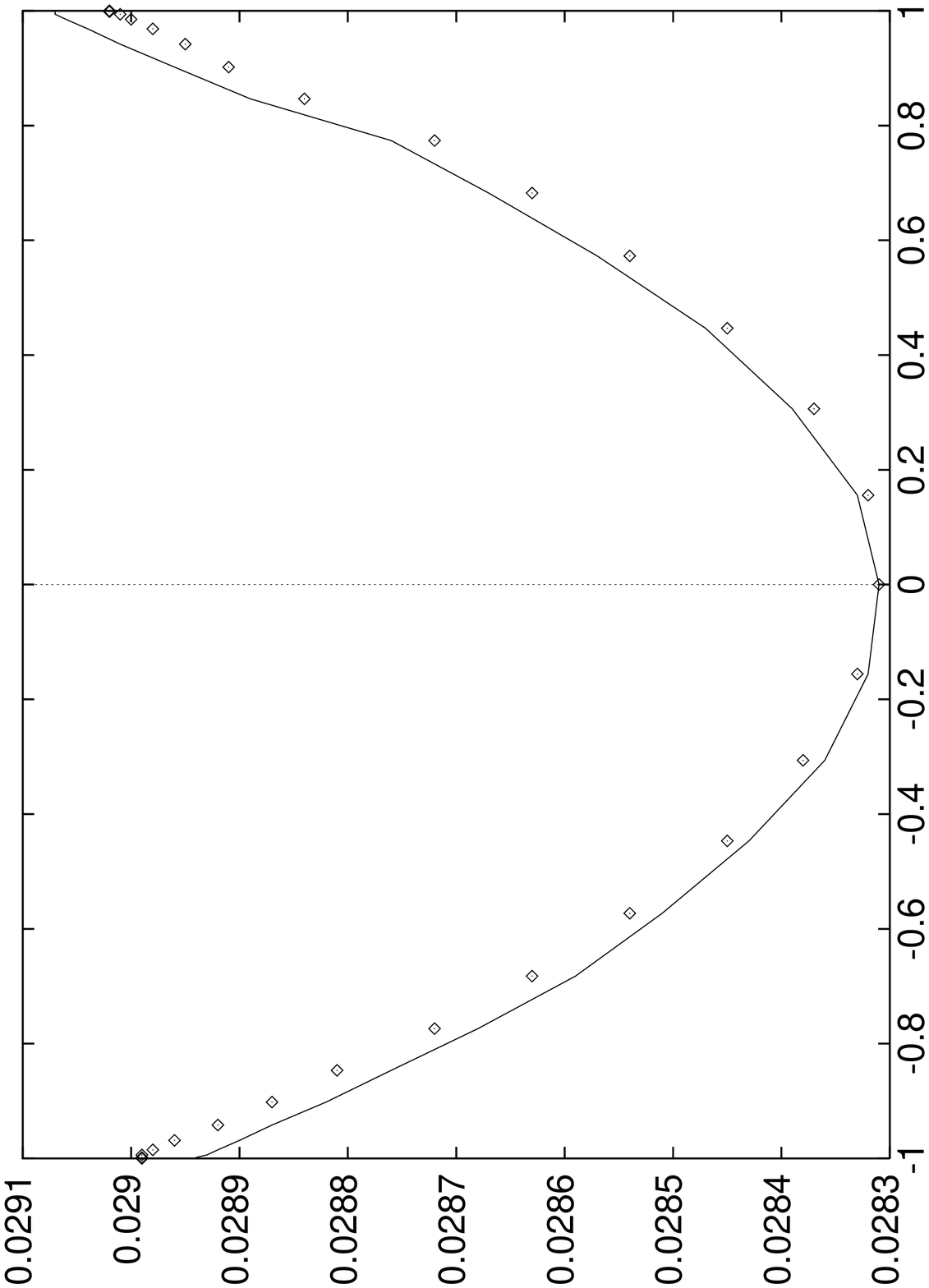}}
\put(310,-120){$ \scriptscriptstyle\cos\Theta_{-}$}
\put(170,0){$ \scriptscriptstyle\frac{d\sigma}{d\Theta_{-}}/pb$}
\label{80w3}
\put(185,-148)
{\parbox{5.7cm}{\scriptsize Fig.~4:
Lepton angular distribution in (C) for $m_0=80$ GeV 
with spin correlations fully taken into account (solid)
and for assumed factorization (dotted).}}
\end{picture}

\vspace{6cm}
\begin{picture}(10,5)
\put(0,0){\includegraphics{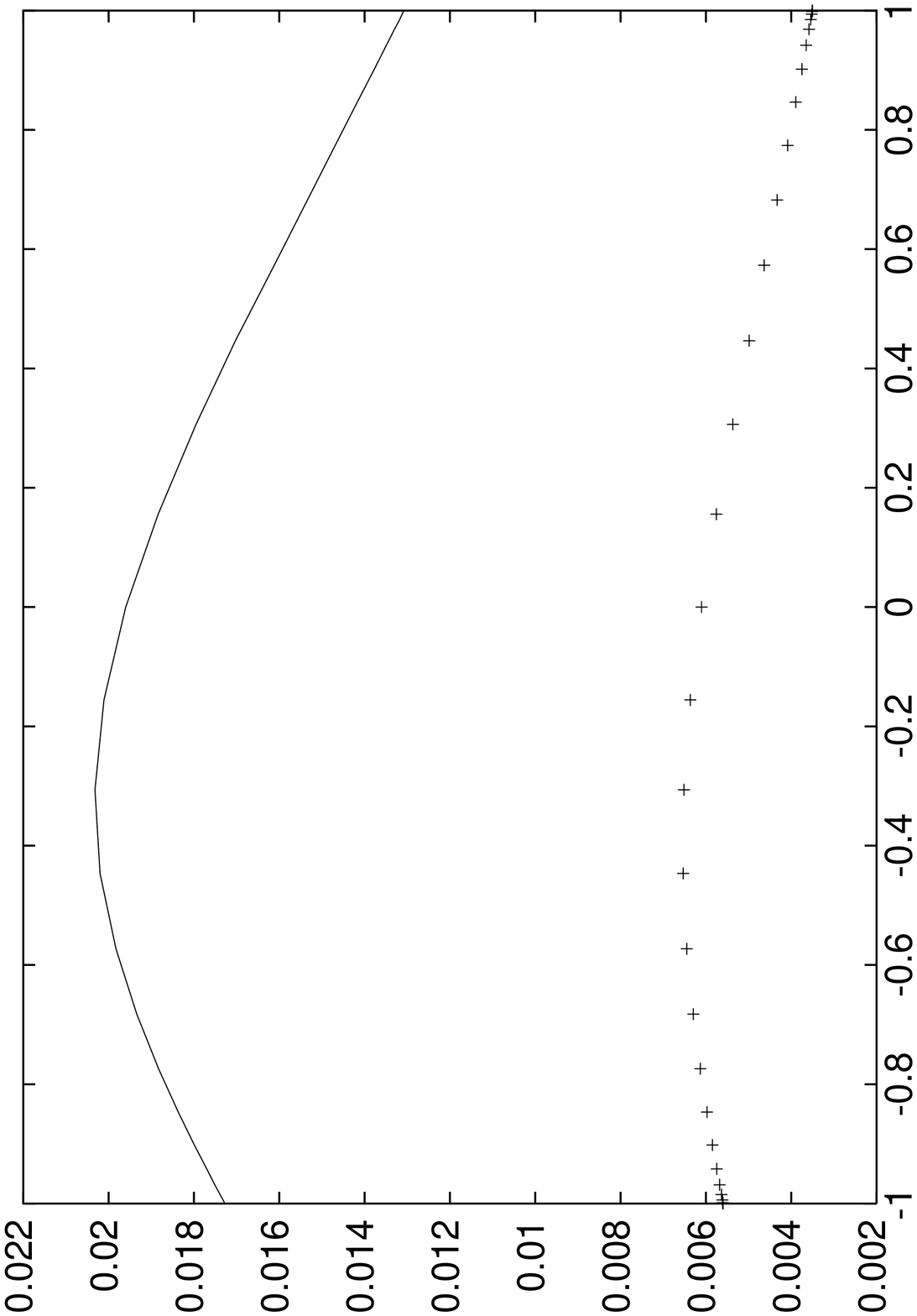}}
\put(130,-120){$ \scriptscriptstyle\cos\Theta_{+-}$}
\put(-10,0){$ \scriptscriptstyle\frac{d\sigma}{d\Theta_{+-}}/pb$}
\put(-10,-156)
{\parbox{5.7cm}{\scriptsize Fig.~5:
Opening angle distribution in (A) for $m_0=80$ GeV
with spin correlations fully taken into account (solid);
 for $m_0=200$ GeV
with spin correlations (dotted).}}
\put(12,0){\includegraphics{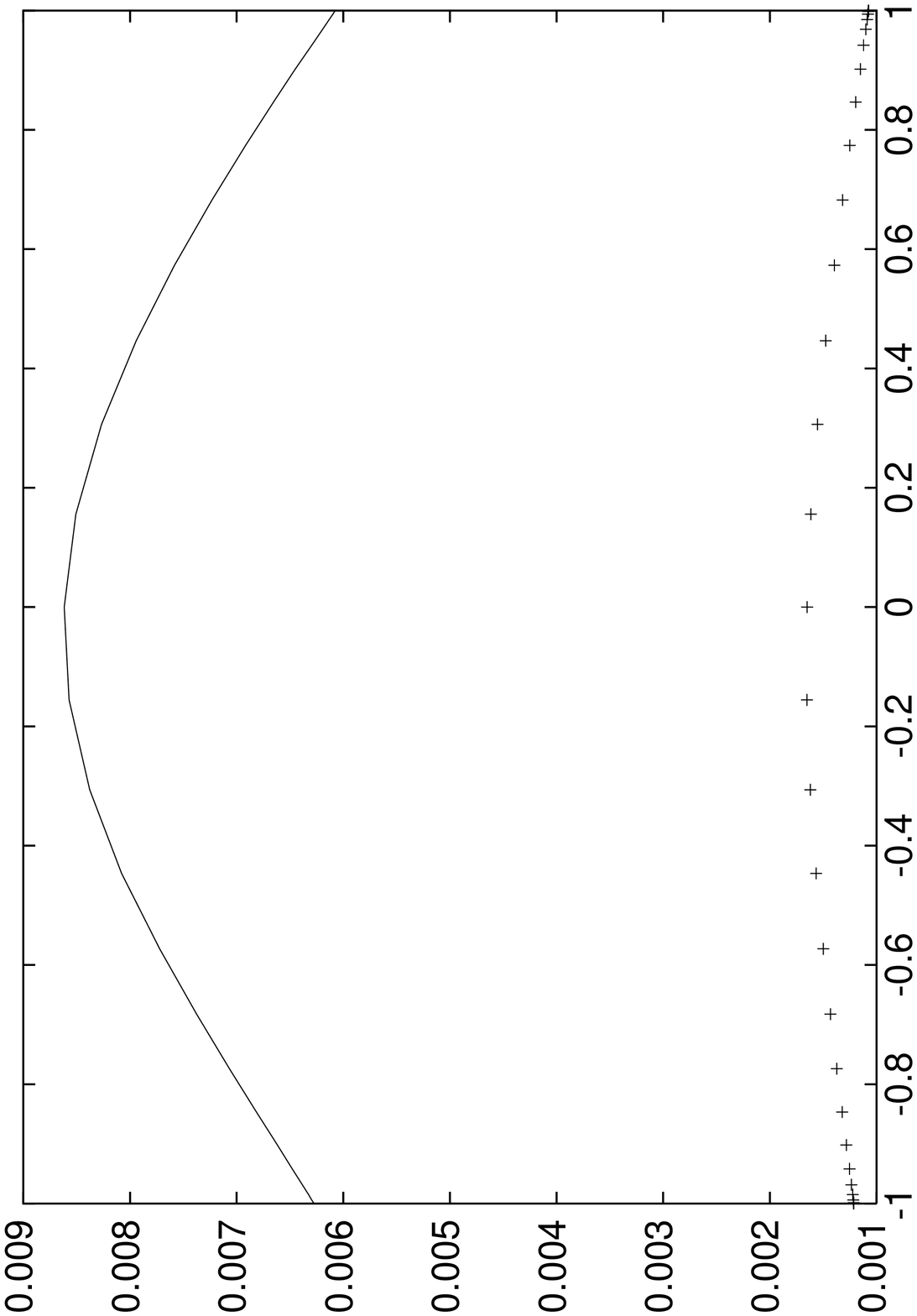}}
\put(310,-120){$ \scriptscriptstyle\cos\Theta_{+-}$}
\put(170,0){$ \scriptscriptstyle\frac{d\sigma}{d\Theta_{+-}}/pb$}
\put(185,-156)
{\parbox{5.7cm}{\scriptsize Fig.~6:
Opening angle distribution  in (B) for $m_0=80$ GeV
with spin correlations fully taken into account (solid);
 for  $m_0=200$ GeV with spin correlations
(dotted).}}
\end{picture}
\label{zw2}

\vspace{6cm}
\begin{picture}(10,5)
\put(0,0){\includegraphics{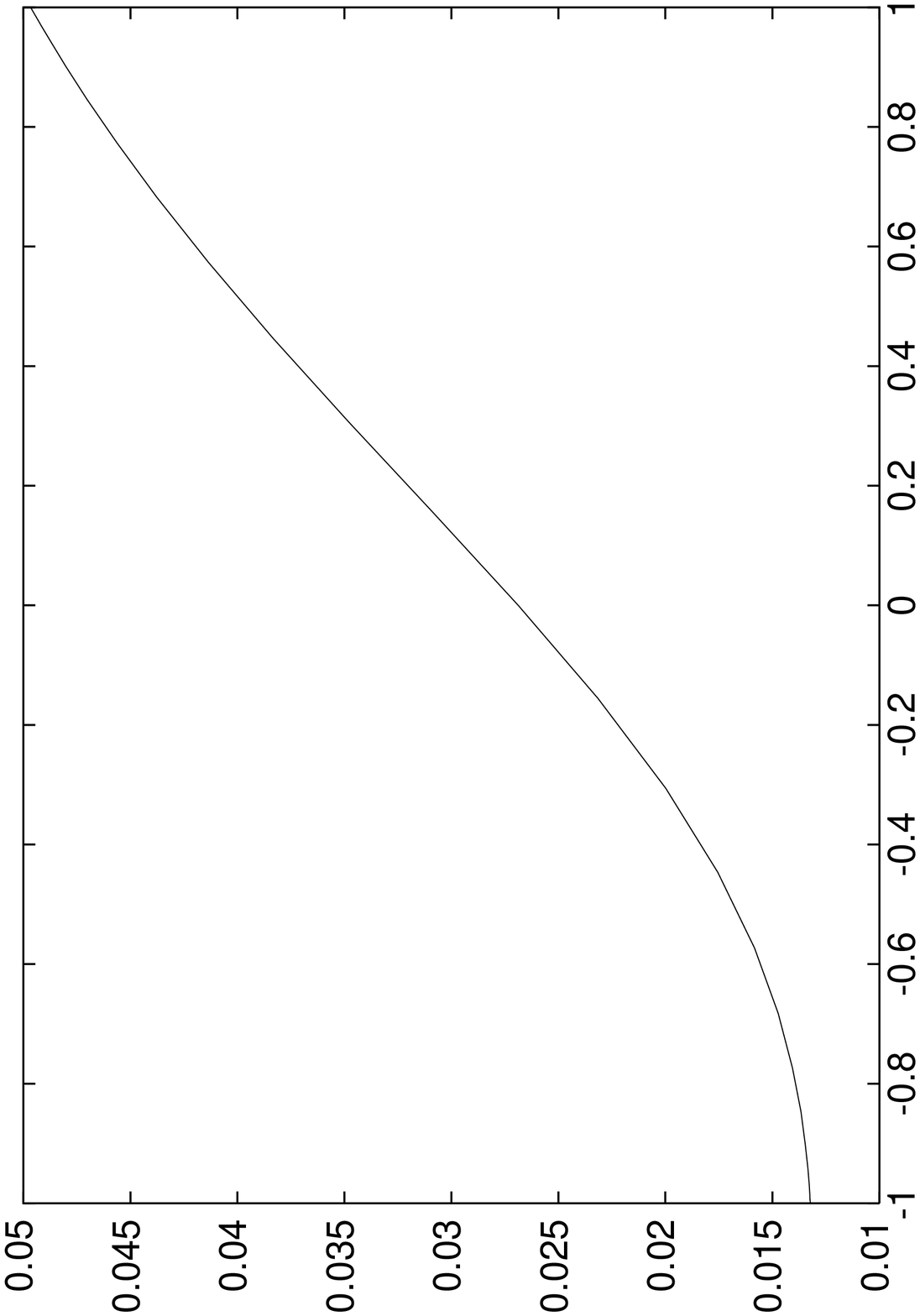}}
\put(130,-120){$ \scriptscriptstyle\cos\Theta_{+-}$}
\put(-10,0){$ \scriptscriptstyle\frac{d\sigma}{d\Theta_{+-}}/pb$}
\label{80zw3}
\put(-10,-140)
{\parbox{5.7cm}{\scriptsize Fig.~7:
Opening angle distribution in (C) for $m_0=80$ GeV
with spin correlations fully taken into account.}}
\put(12,0){\includegraphics{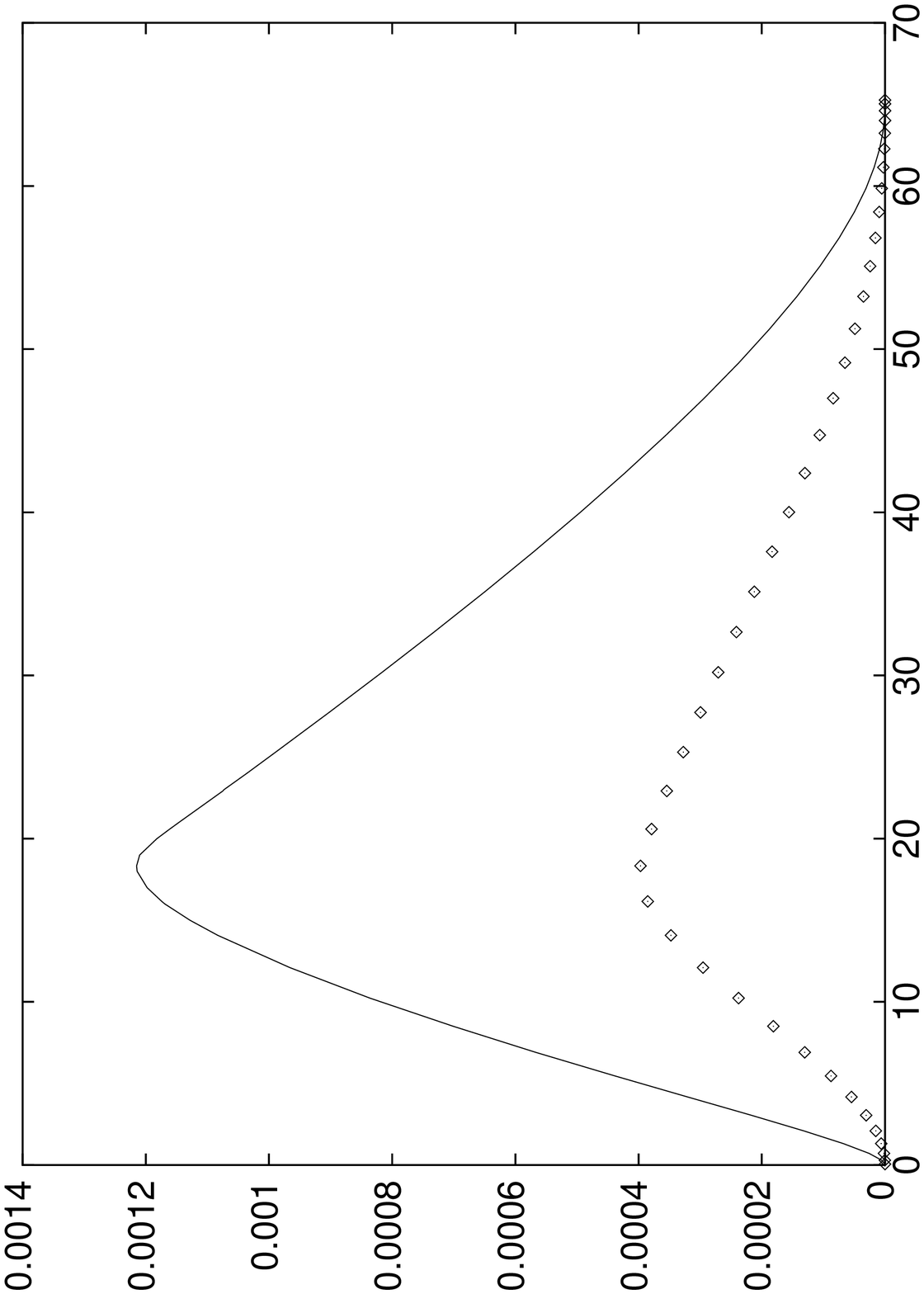}}
\put(310,-120){$ \scriptscriptstyle E_{-}/GeV$}
\put(170,0){$ \scriptscriptstyle \frac{d\sigma}{d E_{-}}$}
\put(185,-140)
{\parbox{5.7cm}{\scriptsize Fig.~8:
Energy distribution in (A) for $m_0=80$ GeV (solid) and
 for $m_0=200$ GeV (dotted).}}
\end{picture}
\label{e1}
\end{document}